\documentstyle[twoside]{article}

\catcode`\@=11
\long\def\@makefntext#1{
\protect\noindent \hbox to 3.2pt {\hskip-.9pt  
$^{{\eightrm\@thefnmark}}$\hfil}#1\hfill}		

\def\thefootnote{\fnsymbol{footnote}}
\def\@makefnmark{\hbox to 0pt{$^{\@thefnmark}$\hss}}	
	
\def\ps@myheadings{\let\@mkboth\@gobbletwo
\def\@oddhead{\hbox{}
\rightmark\hfil\eightrm\thepage}   
\def\@oddfoot{}\def\@evenhead{\eightrm\thepage\hfil
\leftmark\hbox{}}\def\@evenfoot{}
\def\sectionmark##1{}\def\subsectionmark##1{}}

\oddsidemargin=\evensidemargin
\renewcommand{\thefootnote}{\fnsymbol{footnote}}

\newcounter{sectionc}\newcounter{subsectionc}\newcounter{subsubsectionc}
\renewcommand{\section}[1] {\vspace{12pt}\addtocounter{sectionc}{1} 
\setcounter{subsectionc}{0}\setcounter{subsubsectionc}{0}\noindent 
	{\tenbf\thesectionc. #1}\par\vspace{5pt}}
\renewcommand{\subsection}[1] {\vspace{12pt}\addtocounter{subsectionc}{1} 
	\setcounter{subsubsectionc}{0}\noindent 
	{\bf\thesectionc.\thesubsectionc. {\kern1pt \bfit #1}}\par\vspace{5pt}}
\renewcommand{\subsubsection}[1] {\vspace{12pt}\addtocounter{subsubsectionc}{1}
	\noindent{\tenrm\thesectionc.\thesubsectionc.\thesubsubsectionc.
	{\kern1pt \tenit #1}}\par\vspace{5pt}}
\newcommand{\nonumsection}[1] {\vspace{12pt}\noindent{\tenbf #1}
	\par\vspace{5pt}}

\newcounter{appendixc}
\newcounter{subappendixc}[appendixc]
\newcounter{subsubappendixc}[subappendixc]
\renewcommand{\thesubappendixc}{\Alph{appendixc}.\arabic{subappendixc}}
\renewcommand{\thesubsubappendixc}
	{\Alph{appendixc}.\arabic{subappendixc}.\arabic{subsubappendixc}}

\renewcommand{\appendix}[1] {\vspace{12pt}
        \refstepcounter{appendixc}
        \setcounter{figure}{0}
        \setcounter{table}{0}
        \setcounter{lemma}{0}
        \setcounter{theorem}{0}
        \setcounter{corollary}{0}
        \setcounter{definition}{0}
        \setcounter{equation}{0}
        \renewcommand{\thefigure}{\Alph{appendixc}.\arabic{figure}}
        \renewcommand{\thetable}{\Alph{appendixc}.\arabic{table}}
        \renewcommand{\theappendixc}{\Alph{appendixc}}
        \renewcommand{\thelemma}{\Alph{appendixc}.\arabic{lemma}}
        \renewcommand{\thetheorem}{\Alph{appendixc}.\arabic{theorem}}
        \renewcommand{\thedefinition}{\Alph{appendixc}.\arabic{definition}}
        \renewcommand{\thecorollary}{\Alph{appendixc}.\arabic{corollary}}
        \renewcommand{\theequation}{\Alph{appendixc}.\arabic{equation}}
        \noindent{\tenbf Appendix \theappendixc #1}\par\vspace{5pt}}
\newcommand{\subappendix}[1] {\vspace{12pt}
        \refstepcounter{subappendixc}
        \noindent{\bf Appendix \thesubappendixc. {\kern1pt \bfit #1}}
	\par\vspace{5pt}}
\newcommand{\subsubappendix}[1] {\vspace{12pt}
        \refstepcounter{subsubappendixc}
        \noindent{\rm Appendix \thesubsubappendixc. {\kern1pt \tenit #1}}
	\par\vspace{5pt}}

\newcommand{\textlineskip}{\baselineskip=13pt}
\newcommand{\smalllineskip}{\baselineskip=10pt}

\def\eightcirc{
\begin{picture}(0,0)
\put(4.4,1.8){\circle{6.5}}
\end{picture}}
\def\eightcopyright{\eightcirc\kern2.7pt\hbox{\eightrm c}} 

\newcommand{\copyrightheading}[1]
	{\vspace*{-2.5cm}\smalllineskip{\flushleft
	{\footnotesize #1}
	 }}

\def\abstracts#1#2#3{{
	\centering{\begin{minipage}{4.5in}\baselineskip=10pt\footnotesize
	\parindent=0pt #1\par 
	\parindent=15pt #2\par
	\parindent=15pt #3
	\end{minipage}}\par}} 

\newcommand{\bibit}{\nineit}

\renewenvironment{thebibliography}[1]
	{\frenchspacing
	 \ninerm\baselineskip=11pt
	 \begin{list}{\arabic{enumi}.}
	{\usecounter{enumi}\setlength{\parsep}{0pt}
	 \setlength{\leftmargin 12.7pt}{\rightmargin 0pt} 
	 \setlength{\itemsep}{0pt} \settowidth
	{\labelwidth}{#1.}\sloppy}}{\end{list}}

\newcounter{itemlistc}
\newcounter{romanlistc}
\newcounter{alphlistc}
\newcounter{arabiclistc}

\newcommand{\fcaption}[1]{
        \refstepcounter{figure}
        \setbox\@tempboxa = \hbox{\footnotesize Fig.~\thefigure. #1}
        \ifdim \wd\@tempboxa > 5in
           {\begin{center}
        \parbox{5in}{\footnotesize\smalllineskip Fig.~\thefigure. #1}
            \end{center}}
        \else
             {\begin{center}
             {\footnotesize Fig.~\thefigure. #1}
              \end{center}}
        \fi}

\newcommand{\tcaption}[1]{
        \refstepcounter{table}
        \setbox\@tempboxa = \hbox{\footnotesize Table~\thetable. #1}
        \ifdim \wd\@tempboxa > 5in
           {\begin{center}
        \parbox{5in}{\footnotesize\smalllineskip Table~\thetable. #1}
            \end{center}}
        \else
             {\begin{center}
             {\footnotesize Table~\thetable. #1}
              \end{center}}
        \fi}

\def\@citex[#1]#2{\if@filesw\immediate\write\@auxout
	{\string\citation{#2}}\fi
\def\@citea{}\@cite{\@for\@citeb:=#2\do
	{\@citea\def\@citea{,}\@ifundefined
	{b@\@citeb}{{\bf ?}\@warning
	{Citation `\@citeb' on page \thepage \space undefined}}
	{\csname b@\@citeb\endcsname}}}{#1}}

\newif\if@cghi
\def\cite{\@cghitrue\@ifnextchar [{\@tempswatrue
	\@citex}{\@tempswafalse\@citex[]}}
\def\citelow{\@cghifalse\@ifnextchar [{\@tempswatrue
	\@citex}{\@tempswafalse\@citex[]}}
\def\@cite#1#2{{$\null^{#1}$\if@tempswa\typeout
	{IJCGA warning: optional citation argument 
	ignored: `#2'} \fi}}
\newcommand{\citeup}{\cite}

\def\pmb#1{\setbox0=\hbox{#1}
	\kern-.025em\copy0\kern-\wd0
	\kern.05em\copy0\kern-\wd0
	\kern-.025em\raise.0433em\box0}

\def\fnt#1#2{\footnotetext{\kern-.3em
	{$^{\mbox{\scriptsize #1}}$}{#2}}}

\def\fpage#1{\begingroup
\voffset=.3in
\thispagestyle{empty}\begin{table}[b]\centerline{\footnotesize #1}
	\end{table}\endgroup}

\def\runninghead#1#2{\pagestyle{myheadings}
\markboth{{\protect\footnotesize\it{\quad #1}}\hfill}
{\hfill{\protect\footnotesize\it{#2\quad}}}}
\font\tenrm=cmr10
\font\tenit=cmti10 
\font\tenbf=cmbx10
\font\bfit=cmbxti10 at 10pt
\font\ninerm=cmr9
\font\nineit=cmti9

\font\eightrm=cmr8

\def\qed{\hbox{${\vcenter{\vbox{			
   \hrule height 0.4pt\hbox{\vrule width 0.4pt height 6pt
   \kern5pt\vrule width 0.4pt}\hrule height 0.4pt}}}$}}

\renewcommand{\thefootnote}{\fnsymbol{footnote}}	

\newcommand{\be}{\begin{equation}}
\newcommand{\ee}{\end{equation}}
\newcommand{\bea}{\begin{eqnarray}}
\newcommand{\eea}{\end{eqnarray}}
\newcommand{\bn}{\mbox{\boldmath$\nabla$}}
\newcommand{\p}{\partial}
\newcommand{\bE}{{\bf E}}
\newcommand{\bB}{{\bf B}}
\newcommand{\bx}{{\bf x}}
\newcommand{\La}{\Lambda}

\begin{document}

\runninghead{Decay of Supersymmetric Particles} {Decay of 
Supersymmetric Particles}

\normalsize\textlineskip
\thispagestyle{empty}
\setcounter{page}{1}

\copyrightheading{CLNS-00/1710}			

\vspace*{0.88truein}

\fpage{1}
\centerline{\bf STRING WEBS AND THE DECAY OF}
\vspace*{0.035truein}
\centerline{\bf SUPERSYMMETRIC PARTICLES}
\vspace*{0.37truein}
\centerline{\footnotesize PHILIP C. ARGYRES and K. NARAYAN}
\vspace*{0.015truein}
\centerline{\footnotesize\it Newman Laboratory, Cornell University}
\baselineskip=10pt
\centerline{\footnotesize\it Ithaca, NY 14853, USA}

\vspace*{0.21truein}
\abstracts{The spectrum of stable electrically and magnetically charged 
supersymmetric particles can change discontinuously due to the decay of 
these particles as the vacuum on the Coulomb branch is varied.  We show 
that this decay process is well described by semi-classical field 
configurations purely in terms of the low energy effective action on
the Coulomb branch even when it occurs at strong coupling. The resulting 
picture of the stable supersymmetric spectrum is a generalization of the 
``string web'' picture of these states found in string constructions 
of certain theories.}{}{}

\textlineskip			
\vspace*{12pt}			

\noindent
Four dimensional gauge theories with at least eight supersymmetries
have a Coulomb branch of inequivalent vacua in which the low energy
theory has unbroken $U(1)$ gauge factors.  These theories also have
a spectrum of massive charged particles with various electric and
magnetic charges under the $U(1)$'s, lying in supersymmetry
multiplets.  The masses of those in short multiplets---the BPS
states---are related to their charges by the supersymmetry
algebra.\citeup{wo78} The spectrum of possible BPS masses can then be
determined using supersymmetry selections rules.\citeup{sw9407} This,
however, leaves open the question of the existence and multiplicity of
these states.  Furthermore, even if such a state exists in some region
of the Coulomb branch, it may be unstable to decay at curves of
marginal stability (CMS) on the Coulomb branch.\citeup{cv9211} We
propose\citeup{an0101} a solution to the question of the multiplicity 
of BPS states for $N=2$ supersymmetric theories in four dimensions 
just in terms of the low energy effective $U(1)^n$ action on the 
Coulomb branch.

The form of the answer we get coincides with the ``string web''
picture of BPS states developed in the context of the D3-brane
construction of $N=4$ $SU(n)$ superYang-Mills theory and the F-theory
solution to $N=2$ $SU(2)$ gauge theory with fundamental
matter\citeup{s9605}.  Moreover, our solution generalizes these
constructions to arbitrary field theory data (gauge groups, matter
representations, couplings and masses).  The resulting picture is
quite simple: BPS states are represented by string webs on the Coulomb
branch of the theory with one end at the point corresponding to the
vacuum in question (the analog of the 3-brane probe in the F-theory
picture) and the other ends lying on the complex codimension 1
singularities on the Coulomb branch (the analogs of the $(p,q)$
7-branes of the F-theory picture).  The strands of the string web lie
along geodesics in the Coulomb branch metric.  Each strand carries
electric and magnetic charges under the $U(1)^n$ low energy gauge
group: the total charge flowing into the vacuum point determines the
total charge of the BPS state, while only multiples of the charges
determined by the $Sp(2n,Z)$ monodromies 
\pagebreak

\textheight=7.8truein
\setcounter{footnote}{0}
\renewcommand{\thefootnote}{\alph{footnote}}

\noindent
around the codimension 1 singularities are allowed to flow into 
those ends of the web.

Perhaps the most surprising thing about our solution is that it
describes the stability of the monopole and dyon BPS spectrum wholly
in terms of the $U(1)^n$ low energy effective action.  This is
possible because the distance $\Delta X$ from a given vacuum on the
Coulomb branch to a CMS acts as a new low energy scale which can be
made arbitrarily small compared to the strong coupling scale $\Lambda$
of the non-Abelian gauge theory as we approach the CMS.  In particular,
we show that as we approach the CMS the classical low energy
field configuration describing a state which decays across the CMS
will develop two or more widely separated charge centers (which appear
as singularities in the low energy solution); the distance between
these centers varies inversely with $\Delta X$.  In this limit
the details of the microscopic physics become irrelevant for the decay
of a BPS state across a CMS, which is described by a low energy field
configuration with charge centers becoming infinitely separated.

The classical BPS field configuration of the scalar fields in the low
energy $U(1)^n$ theory will have one or more singularities or sources
where scalar field gradients and $U(1)$ field strengths diverge.  Near
these points the low energy description breaks down and should be cut
off by boundary conditions reflecting the matching onto the
microscopic physics of the non-Abelian gauge theory.  We show that
these boundary conditions are determined by the BPS condition.

Several previous discussions\citeup{gkmtz9903,ly9804,rsvv0006} of the
semi-classical description of BPS states in supersymmetric theories
were the starting point for this work.

\section{BPS states near CMS}
\noindent
We restrict our attention to a single charge sector of the low energy
$U(1)^n$ theory on the Coulomb branch.  The question is whether at a
given vacuum there is or is not a one-particle BPS state in that
charge sector.  The mass of a BPS state is $M=|Z|$ where the central
charge $Z$ is the sum of terms proportional to the charges.  This mass
is the minimum mass of any state (BPS or not, single particle or not)
in this charge sector.  A single BPS particle $M$ in this charge sector
is stable or at worst marginally stable
against decay into two (or more) constituent particles, since
by charge conservation $Z = Z_1 + Z_2$, so by the triangle inequality
$M \le M_1 + M_2$.  The CMS are submanifolds of the Coulomb branch
where this inequality is saturated.  As one adiabatically changes the
order parameter on the Coulomb branch from a vacuum on one side of a
CMS to a vacuum on the other, the one particle state $M$ becomes more
and more nearly degenerate with the two particle state $M_1+M_2$.
Suppose $M$ does decay across the CMS, so only the two particle
state $M_1+M_2$ remains in the spectrum once we have crossed the CMS.
Since the transition takes place right at threshold, $M$ will decay
into the two particle state with zero relative momentum.  

Now, it follows from the triangle inequality that the two particle
state $M_1 + M_2$ is not BPS, even for zero relative momentum, except
precisely at the CMS.  Thus there will generically be no BPS force
cancelation between particles $M_1$ and $M_2$, so the zero relative
momentum two particle state is classically approximated by two
spatially infinitely separated one-particle states (to have a static
configuration).  By locality and the adiabatic theorem, the transition
across the CMS of a one particle state to a widely separated, zero
momentum, two particle state should go by way of field configurations
with large spatial overlap, implying that a state decaying just where
its mass reaches the two-particle threshold of its decay products
should have a diverging spatial extent as it approaches the
transition.  This argument gives the basic physics underlying our
approach to decays across CMS: {\em if a BPS state does decay across a
CMS, that decay will be visible semiclassically in the low energy
effective action even if it takes place at strong coupling from a
microscopic point of view.}  The resulting picture is quite intuitive:
a BPS state decaying across a CMS does so by becoming an ever-larger,
more loosely bound state of its eventual decay products.  Once the CMS
is crossed, the bound state ceases to exist, and so, in particular,
there will be no static BPS solutions to the low energy equations of
motion and boundary conditions in this region of the Coulomb branch.

\section{A $U(1)$ toy example}
\noindent
We illustrate these properties in a simple model which captures
the main physics of the CMS.   Consider the $U(1)$ effective
action with two real scalars $X$, $Y$:
\be\label{toym}
S = -\int d^4 x \left( {1\over4} F_{\mu\nu} F^{\mu\nu}
+ {1\over2} \p_\mu X \p^\mu X + {1\over2} \p_\mu Y \p^\mu Y \right)
\ee
This theory by itself is too simple to be interesting, but all we
need to add to it to capture the essential physics of the decay of BPS
states across CMS are the presence of singularities in the vacuum
manifold.  In this case the vacuum manifold (Coulomb branch) is the
$X$-$Y$ plane.  For concreteness we posit that there are two
singularities at points on the Coulomb branch with coordinates
$(X,Y)=(L,0)$ where a particle with electric charge $(Q_E,Q_B)=(1,0)$
becomes massless, and $(X,Y)=(0,L)$ where a particle with magnetic
charge $(Q_E,Q_B)=(0,1)$ becomes massless.

The low energy effective action (\ref{toym}) includes only the first
terms in a derivative expansion with higher terms suppressed by powers
of a cutoff energy scale $\Lambda$, or distance scale $r_\La \equiv
{1\over\La}$.  Our strategy will be to use the low energy $U(1)$
solution away from the charge cores ($r>r_\La$) and impose appropriate
boundary conditions in the vicinity of the cores ($r\sim r_\La$).  The
qualitative features of these boundary conditions are easy to deduce.
The basic boundary condition on the electric and magnetic fields
follow from Gauss' law, $\oint \bE\cdot d{\bf a} = 4\pi Q_E$, and
$\oint \bB\cdot d{\bf a} = 4\pi Q_B$, where the integrals are over
spheres enclosing the charge cores.

The $X$, $Y$ scalar fields tend to approach their values at Coulomb
branch singularities near the charge cores.  Indeed, such field
configurations are energetically favored because they approach those
of the charged BPS states which become massless at the singularities
$(X,Y) = (0,L)$ or $(L,0)$.  Thus the scalar fields
satisfy the approximate Dirichlet boundary conditions
\be\label{fuzzydirbc}
(X,Y) \simeq (0,L)\ \mbox{or}\ (L,0)\ \ \mbox{within}\ \ B^3_\La 
\ee
where $B^3_\La$ is a ball of approximate radius $r_\La$ around each
charge core and $(X,Y)\simeq(L,0)$ means that $(X,Y)$ pass within a
distance $\La$ of $(L,0)$ on the Coulomb branch.  This ``fuzzy ball''
boundary condition is all we can physically demand of the low
energy solution since it is not accurate on spatial resolutions less
than $r_\La$ nor for field value resolutions less than $\La$.  In the
limit as the vacuum approaches a CMS, as we have described
qualitatively in the last section and will see explicitly below, the
size of the field configuration grows, and so the relative size of the
cutoff region $r_\La$ to the field configuration shrinks.  In
this limit the fuzzy ball Dirichlet boundary conditions become
precise.

To simplify the algebra and make the basic points clear we choose the
vacuum to be symmetrically placed at the point $(X_0,X_0)$ on the
Coulomb branch and choose the state to have electric and magnetic
charges $(Q_E,Q_B)=(1,1)$.  There will then be separate electric and
magnetic charge cores.  The usual BPS equations for a static field
configuration following from (\ref{toym}) are then simply
$\nabla^2 X = \nabla^2 Y = 0$
while the electric and magnetic fields are determined in terms
of $X$ and $Y$ by $\bE = \bn X$, and $\bB = \bn Y$.
The Gauss constraint implies a solution of the 
approximate form
\be
X - X_0 = {1\over |\bx-\bx_E|},\qquad Y - X_0 = {1\over |\bx-\bx_B|}
\ee
so that the electric and magnetic charge cores are located at $\bx_E$
and $\bx_B$ respectively.  We want to check whether there are values
of $\bx_E$ and $\bx_B$ for which this solution satisfies the fuzzy
ball Dirichlet boundary condition (\ref{fuzzydirbc}), which demands
that the $(X,Y)$ values taken by a solution enters a small ball around
$(L,0)$ as $\bx\to\bx_E$ (and a small ball around $(0,L)$ as
$\bx\to\bx_B$).

Now for $|\bx-\bx_E| \simeq (L-X_0)^{-1}$ it follows that $X \simeq L$
as desired near the electric singularity.  But this then implies that
$Y \simeq X_0 + r_{EB}^{-1}$ where $r_{EB}\equiv|\bx_E-\bx_B|$ is the
spatial distance between the charge centers.  The fuzzy Dirichlet
boundary condition $Y\simeq 0$ then implies that the separation of
the charge centers must be fixed to be
\be
r_{EB} = -{1\over X_0} .
\ee
There are two striking things about this conclusion.  First, there is
only a solution for $X_0<0$.  This inequality is equivalent to saying
that there is only a solution on one side of the CMS for our dyonic
$(Q_E,Q_B)=(1,1)$ state (as can be checked from the explicit BPS mass
formula).  Second, as $X_0\to 0^-$ ({\it i.e.}\ as we approach the
CMS) the spatial separation of the electric and magnetic charge
centers goes to infinity.  This confirms the qualitative description
of the decay of a supersymmetric state near the CMS given above.

As we describe in detail elsewhere\citeup{an0101}, in the limit as the
vacuum approaches the CMS, the projection on the Coulomb branch of the
above low energy solution more and more accurately approximates a
string web, {\it i.e.}\ it degenerates to a collection of segments of
geodesics---``strings''---on the Coulomb branch which can meet in
3-string junctions or end at the vacuum or singularities on the Coulomb
branch.  For example, in the $U(n)$ $N=4$ SYM theory these ``string''
webs are thus stretched in the $6n$-dimensional moduli space and end
on $6(n-1)$-dimensional singular submanifolds.  Note that this is
different from the string web picture found in the string construction
of this theory, where the strings are stretched in the 6-dimensional
space transverse to $N$ D3-branes.  It can be shown\citeup{an0101} that
the 6-dimensional string theory webs are obtained from our
$6n$-dimensional webs by a simple mapping.  The basic reason this
works is that the $6N$-dimensional moduli space ${\cal M}$ of the
$U(n)$ theory is ${\cal M} = (R^6)^n/S_n$ where the permutation group
$S_n$ interchanging the $R^6$ factors is the Weyl group of $U(n)$.

Another property of the string webs found in string constructions is
that 3-pronged string states that stretch between D-branes satisfy
charge conservation and tension balance.  These properties are also
reproduced\citeup{an0101} in the webs from our low energy solutions.
However, the low energy solutions have an extended spatial structure
which is not seen in the string constructions.  For example, the
configuration discussed above corresponds to a $(1,0)$ (fundamental
string), a $(0,1)$ (D-string)and a $(1,1)$ string stretched between
three D3-branes.  The three prongs or string legs meet at a common
point, the junction, which remains point-like even arbitrarily close
to the CMS. In the field theory picture discussed here, we
have seen that the leg corresponding to the decaying $(1,1)$ dyon
grows in spatial size ({\em i.e.}, along the D3-brane worldvolume)
near the CMS. Arbitrarily close to the CMS, the configuration
resembles two separate strings that end on the D3-brane corresponding
to the decaying $U(1)$. Thus there does not appear to be a point-like
junction as the decaying configuration approaches the CMS in the field
theory picture.

The difference between the two perspectives is the result of different
orders of limits.  Consider the Dirac-Born-Infeld action for the
D3-brane corresponding to the decaying dyon, treating it as a probe in a
background of other D3-branes. The brane prong corresponding to the
decaying $(1,1)$ dyon then ends on other D3-branes that are treated as a
fixed background.  Looking at the quadratic terms and comparing with the
low energy effective action we have used above, we can see that the
scalars $X$ in the field theory with mass dimension unity and the
coordinates in the brane transverse space, $x$, are related by $X =
x/\alpha'$, where $\alpha'$ is the string length squared. The separation
between the charge centers in the D-brane worldvolume is
\be
r_{EB} \sim 1/(-X_0) =  \alpha' / (-x_0).
\ee
The low energy field theory is a good approximation in the $\alpha' \to
0$ limit holding the scalar vevs, including $X_0$, fixed.  On the other
hand, in the string junction/geodesic picture in string theory, the
coordinates $x_0$ are what are held fixed: taking $\alpha' \to 0$ to
suppress higher stringy corrections gives a vanishing separation
$r_{EB}$ between the charge centers, {\em i.e.}\ a pointlike junction.  
For $x_0 = 0$, the separation is indeterminate, which corresponds to the
two strings ending anywhere on the D-brane.  This recovers the string
theory result and it further illustrates that a {\em spatially}\ 
point-like junction is not visible within the field theory approximation.

\nonumsection{Acknowledgments}
\noindent
It is a pleasure to thank A.~Buchel, J.~Hein, R.~Maimon,
J.~Maldacena, M.~Moriconi, S.~Pelland, M.~Rangamani, V.~Sahakian, and 
A.~Shapere for
helpful comments and discussions.  This work was supported in part by NSF
grant PHY95-13717.

\nonumsection{References}

\end{document}